\documentclass[conference]{IEEEtran}

\usepackage{cite}
\usepackage{amsmath,amssymb,amsfonts}
\usepackage{algorithmic}
\usepackage{graphicx}
\usepackage{textcomp}
\usepackage{xcolor}
\usepackage{comment}
\usepackage{csquotes}
\usepackage{dirtytalk}
\usepackage{booktabs}
\usepackage{multirow}
\usepackage{tabularx}
\usepackage{adjustbox}
\usepackage{flushend}
\usepackage{lipsum}
\graphicspath{ {./images/} }

\def\BibTeX{{\rm B\kern-.05em{\sc i\kern-.025em b}\kern-.08em
    T\kern-.1667em\lower.7ex\hbox{E}\kern-.125emX}}
\begin{document}

\title{Exploring Requirements Elicitation from App Store User Reviews Using Large Language Models}

\author{\IEEEauthorblockN{Tanmai Kumar Ghosh}
\IEEEauthorblockA{\textit{Department of Computer Science} \\
\textit{Boise State University}\\
Boise, Idaho, USA \\
tanmaighosh@u.boisestate.edu}
\and
\IEEEauthorblockN{Atharva Pargaonkar}
\IEEEauthorblockA{\textit{Department of Computer Science} \\
\textit{Boise State University}\\
Boise, Idaho, USA \\
atharvapargaonka@u.boisestate.edu}
\and
\IEEEauthorblockN{Nasir U. Eisty}
\IEEEauthorblockA{\textit{Department of Computer Science} \\
\textit{Boise State university}\\
Boise, Idaho, USA \\
nasireisty@boisestate.edu}
}

\maketitle

\begin{abstract}
Mobile applications have become indispensable companions in our daily lives. Spanning over the categories from communication and entertainment to healthcare and finance, these applications have been influential in every aspect. Despite their omnipresence, developing apps that meet user needs and expectations still remains a challenge. Traditional requirements elicitation methods like user interviews can be time-consuming and suffer from limited scope and subjectivity. This research introduces an approach leveraging the power of Large Language Models (LLMs) to analyze user reviews for automated requirements elicitation. We fine-tuned three well-established LLMs BERT, DistilBERT, and GEMMA, on a dataset of app reviews labeled for usefulness. Our evaluation revealed BERT's superior performance, achieving an accuracy of 92.40\% and an F1-score of 92.39\%, demonstrating its effectiveness in accurately classifying useful reviews. While GEMMA displayed a lower overall performance, it excelled in recall (93.39\%), indicating its potential for capturing a comprehensive set of valuable user insights. These findings suggest that LLMs offer a promising avenue for streamlining requirements elicitation in mobile app development, leading to the creation of more user-centric and successful applications.
\end{abstract}


\section{Introduction}
In the dynamic sphere of software development, accurately and efficiently eliciting user requirements~\cite{garcia2020improving} is the pinnacle of the success of any application and the corresponding developers. The rapid growth of mobile apps, characterized by an ever-increasing variety of user preferences and expectations, presents a significant challenge for developers. Traditional requirements elicitation methods like Interview Based Elicitation~\cite{ferrari2022requirements, patkar2020caveats} are effective to a certain extent. While these methods have served as foundational approaches to requirements elicitation, they face limitations in handling the sheer volume and complexity of user-generated content in app-store environments. The reliance on manual analysis and interpretation of reviews can be labor-intensive, time-consuming, and prone to subjective biases. Moreover, the scalability of traditional methods becomes challenging~\cite{8513829} as the number of apps and users grows exponentially across a variety of domains.

In recent years, the emergence of LLMs has revolutionized various natural language processing (NLP) tasks, offering unprecedented capabilities in understanding and generating human-like text. This technology presents a promising approach to streamlining the requirements elicitation process, enhancing accuracy, and extracting valuable insights from the user feedback available. Pre-trained LLMs have the ability to comprehend nuanced language patterns, discern user sentiments, and extract key insights from diverse sources of user feedback. This capability accelerates the requirements elicitation process and enables developers to gain deeper insights into user needs and preferences. It also facilitates more informed decision-making throughout the app development lifecycle.

The integration of LLMs into the requirements elicitation process offers a useful approach to address the critical challenge of fake reviews~\cite{martens2019towards}. LLMs have the ability to analyze linguistic cues, contextual information, and patterns of engagement within user-generated content. It can leverage the effective identification of suspicious or fraudulent reviews. The detection and mitigation of these fake reviews will help maintain the integrity of app-store ecosystems. By excluding misleading and potentially harmful content from the RE process, LLMs can help guarantee that app development reflects genuine user needs and preferences.

Platforms like Google Play Store and Apple App Store have evolved to host the applications. In these platforms, users have the freedom to share their experiences with the application in the form of ratings and written reviews. This action makes the interface dynamic and a symbiotic process between the developers and users. The developers can dive into the insights from the meaningful reviews and improve the application using them as the requirements. There is a big question here: Is every review considerable? Our answer is a humble NO. Hence, it is vital to detect false reviews to elicit insightful requirements.

This research explores the potential of automated requirements elicitation from app store user reviews utilizing LLMs. We focus specifically on leveraging the capabilities of LLMs to analyze user reviews, a source of unstructured data alongside valuable insights into user preferences, expectations, and pain points. By employing LLMs, we aim not only to automate the extraction of requirements from user reviews but also to detect anomalies, such as not useful reviews (reviews having no insight), that may hinder the elicitation process~\cite{lee2023determining, genc2017systematic}.


In the following sections, we will describe the strategies employed in adopting LLMs for requirements elicitation, explore the benefits and potential drawbacks of this approach, and investigate its comparative advantages over traditional methods. Additionally, we will examine the challenges inherent in leveraging LLMs for requirements elicitation and propose direction for future research and development in this domain.

\section{Related Works}

\subsection{Traditional Requirements Elicitation}

Various methods can be used to gather requirements from stakeholders, such as interviews, focus groups, workshops, and questionnaires~\cite{dieste2010systematic}. In recent years, App store-inspired requirements elicitation has gained popularity. Ferrari et al.~\cite{ferrari2023strategies} studied the strategies, benefits, and challenges of App Store-inspired Elicitation (ASE) and made a quantitative comparison of the ASE with Interview Based Elicitation (IBE). They validated their approach with the 58 analysts who were given task questionnaires individually and isolatedly. 

Data-driven requirements engineering (DDRE) is a relatively new approach that can be used to develop and manage requirements from large datasets like user reviews. The majority of research conducted on DDRE has been centered around evaluating various techniques of machine learning, information retrieval, and natural language processing for the purpose of automated requirements mining and classification~\cite{nakamura2021requirements}. However, understanding the usage of DDRE from the developer's perspective, whether they will use the system or not, is required. Nakamura et al.~\cite{nakamura2021requirements} outline a method for data-driven requirements elicitation that utilizes text mining techniques to extract user reviews from app stores. The approach provides a visual representation of the frequently used terms, along with their associated reviews. The study's key results suggest that practitioners received it positively. They found it convenient to use and valuable for identifying requirements while expressing a willingness to incorporate it into their requirements elicitation process.

Manual analysis of the Apple App Store and Google Play Store is a difficult task due to the large number of apps and associated data, making it impractical and labor-intensive. Researchers in the RE community use different techniques to extract data from app stores in order to bridge this gap. Most current methods focus on updating an application and assume that it can be improved by comparing it with similar ones~\cite{canedo2020application}. However, minimal research has been carried out on the first design phase of an application. The insightful work undertaken by Wei~\cite{wei2023enhancing} aims to develop valuable tools that can facilitate app store mining, significantly reducing the workload of requirements engineers and streamlining the requirements elicitation process. They have made significant progress in developing Mini-BAR, a powerful tool to extract valuable insights from app reviews. This innovative tool enables users to classify, cluster, summarize easily, and rank reviews, providing a comprehensive understanding of user feedback. Mini-BAR has already proven to be a valuable asset for developers seeking to improve their apps and better serve their customers. 

It's worth mentioning that a group of researchers has developed a valuable tool called UI-Diffuser~\cite{wei2023boosting}, which can significantly enhance the process of GUI prototyping and aid in refining UI-related requirements. In addition, they aim to create a robust knowledge repository by leveraging the data gathered from these procedures. The end objective is to offer valuable insights for requirements engineers, thereby elevating the efficiency of requirements elicitation.

\subsection{Machine Learning in Requirements Elicitation}
An evolving practice in requirements gathering involves the application of machine learning (ML) methods to streamline the demanding process of requirement management.

Aslam et al.~\cite{aslam2020convolutional} have proposed a remarkable deep-learning approach for app review classification. It employs a deep learning technique that has proved to be more accurate for text classification in various domains. The proposed approach extracts both textual and non-textual information about each app review, preprocesses the textual data, computes the sentiment of app reviews using Senti4SD, and determines the reviewer's history, including the total number of reviews posted by the reviewer and their submission rate (that is, what percentage of their reviews have been submitted for the associated app). The approach involves creating a digital vector for each app review, followed by training a deep learning-based multi-class classifier to classify the reviews. The proposed approach has been evaluated on a public dataset, and the results show that it significantly enhances the state of the art. 

Lately, a lot of attention has been paid to analyzing and organizing feedback in order to generate helpful software maintenance requests. This includes identifying and categorizing bug reports and requests for new features. While there is a lot of focus on gathering and analyzing user reviews, there seems to be a lack of attention given to the Non-Functional Requirements (NFRs) that are expressed in these reviews. Jha et al.~\cite{jha2019mining}  believe there is a lot of value in extracting and synthesizing these NFRs, which can help understand the overall user experience. 

Striving to satisfy user requirements is a crucial factor in achieving this goal and increasing overall user satisfaction. The authors of the paper~\cite{jha2019mining} conducted a two-phase study to extract NFRs from user reviews that were accessible on mobile app stores. In the first phase, a qualitative analysis was conducted using a dataset of 6,000 user reviews that were sampled from various iOS app categories. This analysis aimed to identify the common issues that users faced while using the apps and to gain insights into the areas that needed improvement. According to the findings of the research, it has been observed that 40\% of the reviews analyzed in the dataset highlight the presence of NFRs. To assist developers in addressing these NFRs, researchers have developed an optimized dictionary-based multi-label classification approach to capture them in user reviews automatically. The proposed approach has been evaluated over a dataset of 1,100 user reviews from iOS and Android apps, delivering promising results.
This method could be advantageous for developers to understand user feedback better and improve the quality and user experience of their mobile applications.

\subsection{Generative Adversarial Network (GAN) in Requirements Elicitation}

Natural Language Processing for Requirements Engineering (NLP4RE) is aimed at enhancing the quality of requirements in the RE process by using NLP tools and techniques. However, there has been limited research on utilizing Generative AI-based NLP tools and techniques for requirements elicitation. Recently, LLMs, such as ChatGPT, have gained significant recognition for their outstanding performance in NLP tasks~\cite{ronanki2023investigating}. To explore the potential of ChatGPT in assisting requirements elicitation processes, a recent study~\cite{ronanki2023investigating} formulated six questions to elicit requirements using ChatGPT and conducted interview-based surveys with five RE experts from academia and industry, collecting 30 responses containing requirements. The study evaluated the quality of these 36 responses (human-formulated + ChatGPT-generated) over seven different requirements quality attributes and conducted a second round of interview-based surveys with another five RE experts to compare the quality of requirements generated by ChatGPT with those formulated by human experts. The study found that ChatGPT-generated requirements are highly Abstract, Atomic, Consistent, Correct, and Understandable. Based on these results, the study identifies the most pressing issues related to LLMs and suggests future research that should focus on leveraging the emergent behavior of LLMs more effectively in natural language-based RE activities. These findings hold great promise for improving the quality of requirements and streamlining the RE process.

However, the exploration of not useful app review detection within the context of app store reviews remains less explored area in the  literature on Requirements Engineering.

\section{Data Collection}

\subsection{Dataset Description}
This section describes our dataset used to train and evaluate LLMs in identifying useful app reviews for developers. The dataset focuses on user reviews from the Google Play Store, aiming to assess the informativeness of user feedback for future app development and updates. 
    
We constructed the dataset by collecting reviews from the Google Play Store using web scraping techniques. It comprises over 3,200 reviews, ensuring a sufficient sample size for robust LLM training. Notably, we took care to create a well-balanced dataset. It contains approximately 1,600 reviews labeled as ``useful" and 1,600 labeled as ``not useful" for developers. This balanced distribution mitigates potential biases towards positive or negative reviews, allowing the LLMs to learn to identify valuable insights regardless of the overall review tone.

\begin{figure}[h]
    \includegraphics[width=8cm]{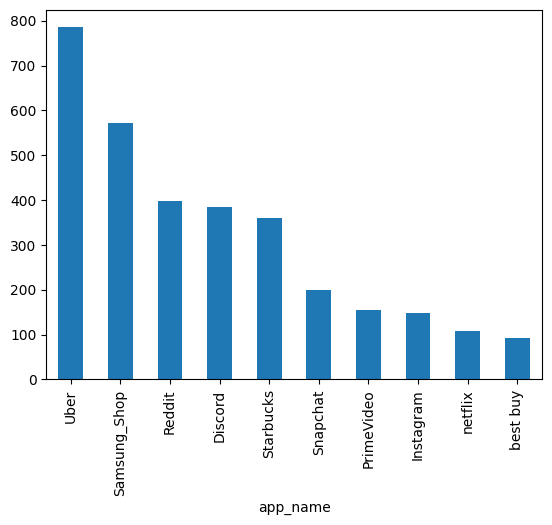}
    \caption{App Reviews distribution}
    \label{fig1}
\end{figure}

\subsection{Dataset Structure}

Each review within the dataset includes the following attributes:

\begin{itemize}
    \item \textbf{AppName:} Name of the application the review pertains to.
    \item \textbf{Username:} Username of the reviewer.
    \item \textbf{app\_rating\_given:} Numerical rating from 1 - 5 provided by the reviewer.
    \item \textbf{review\_description:} Textual content of the review describing the issue.
    \item \textbf{target\_variable:} Categorical label assigned manually by us indicating whether the review is considered \textit{``useful"} or\textit{ ``not useful"} for informing future app development decisions.
\end{itemize}

\subsection{Defining Useful vs. Not Useful Reviews}

To analyze the user reviews for developers, we defined the target variables ``useful" or ``not useful" based on the level of actionable information a review provides. Reviews classified as ``useful" contained specific insights for developers, such as bug reports or feature suggestions. Conversely, reviews categorized as ``not useful" included generic praise, emotional outbursts, or irrelevant comments. 

\begin{figure*}
  \includegraphics[width=18cm,height=9cm]{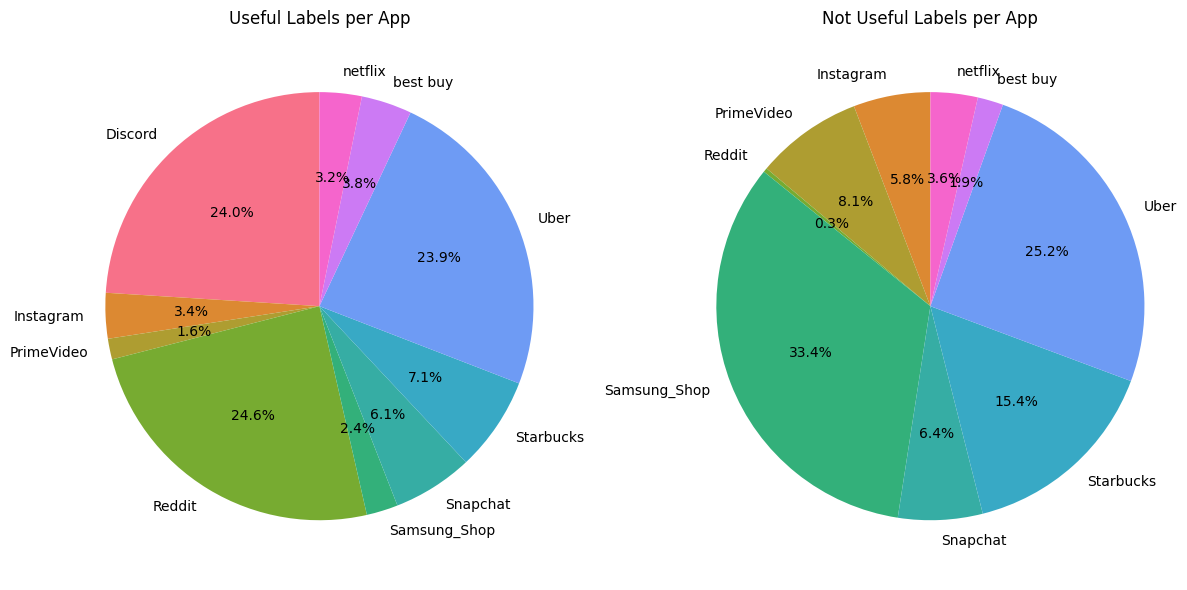}
  \label{useful_vs_not_useful}
  \caption{Distribution of Useful and Non-Useful labels across Apps}
  \label{useful_vs_not_useful}
\end{figure*}


We also considered a wide range of applications from different categories for labeling the reviews, which is shown in Fig~\ref{fig1} and \ref{useful_vs_not_useful}.

By manually assigning these labels, we ensured the target variable was directly aligned with the goal of requirements elicitation. This allows the LLMs to learn to identify reviews containing valuable insights that developers can use to work on improving their apps.

\section{Model Description}
LLMs have emerged as powerful tools for various NLP tasks, including text classification. In this paper, we made use of the performance of three well-established LLMs, BERT, DistilBERT, and GEMMA (recently introduced by Google), to classify user reviews taken from the app store.

\subsection{BERT}

BERT (Bidirectional Encoder Representations from Transformers)~\cite{devlin2018bert}, BERT is a pre-trained deep learning model based on transformer architecture. It utilizes a masked language modeling (MLM) objective function, where the model predicts masked words based on the surrounding context. This pre-training on a massive corpus of text allows BERT to capture rich contextual representations of words and their relationships within the text, making it well-suited for various downstream NLP tasks like text classification.

\subsection{DistilBERT}

DistilBERT is a language model that is relatively smaller in terms of parameters and a faster version of BERT~\cite{sanh2020distilbert}. DistilBERT achieves this by leveraging knowledge distillation, where a smaller student model learns from a larger, pre-trained model. DistilBERT retains a significant portion of the performance of BERT while offering a more efficient alternative for tasks where computational resources are limited.

\subsection{GEMMA}

Google's Gemma presents a significant contribution to the open-source NLP landscape. Inspired by the Gemini models, Gemma offers two parameter sizes (2B and 7B) designed explicitly for efficient deployment on various hardware, including consumer-grade GPUs and CPUs~\cite{gemmateam2024gemma}. This accessibility empowers researchers and developers to experiment with LLMs without requiring high-performance computing resources~\cite{gemmateam2024gemma}.

A comparison of the model characteristics and features is presented in Table~\ref{tab:llms_comparison}.

\begin{table}[htbp]
    \caption{Comparison of Features in Language Models}
    \begin{center}
    \begin{tabular}{|c|p{1cm}|p{2cm}|p{1.5cm}|p{1.5cm}|}
        \hline
        \textbf{Features} & \textbf{GEMMA} & \textbf{BERT} & \textbf{DistilBERT} \\ 
        \hline
        \textbf{Developer} & Google AI & Google AI & Hugging Face \\
        \hline
        \textbf{Type of Models} & Encoder-decoder & Encoder-only & Text-to-text \\
        \hline
        \textbf{Focus} & Generation & General purpose NLP tasks & Lightweight, efficient version of BERT \\
        \hline
        \textbf{Parameters} & Varies (e.g., GEMMA-2B: 2 billion parameters) & 100M - 137B parameters (depending on the version) & 66M parameters (distilled from BERT-base) \\
        \hline
    \end{tabular}
    \end{center}
    \label{tab:llms_comparison}
\end{table}

\section{Study Design}
\begin{figure}[h]
\includegraphics[width=8cm]{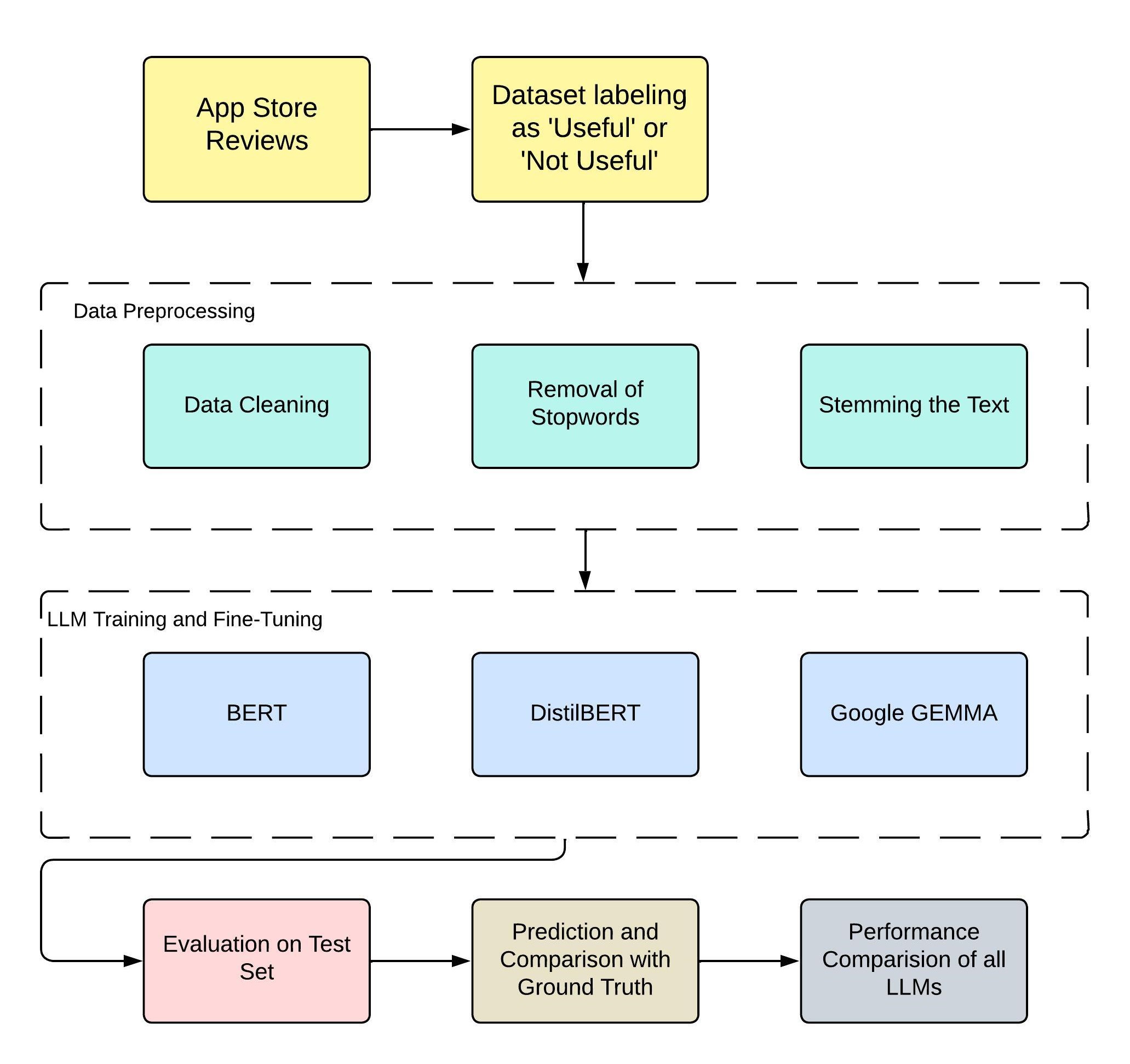}

\caption{Process Diagram}
\label{study_design}
\end{figure}

This study presents a novel approach that leverages the power of LLMs to automate the identification of useful app reviews from a mobile app development perspective. Our focus here lies on reviews collected from the Google Play Store, aiming to bridge the gap between user voices and the future direction of app development. Fig~\ref{study_design} illustrates the summary of the processes performed in our research.

\subsection{Data Pre-Processing}

To prepare the reviews for LLM analysis, we employed a comprehensive data preprocessing pipeline utilizing functionalities from established libraries. This pipeline addressed several key aspects:

\begin{itemize}
    \item \textbf{Text Normalization}: All text was converted to lowercase for consistency. This ensures the LLM focuses on the meaning of words rather than capitalization variations.
    \item \textbf{Data Cleaning}: Special characters, HTML tags, and URLs were meticulously removed using regular expressions. This eliminates irrelevant information that could potentially mislead the LLM's analysis.
    \item \textbf{Tokenization}: Each review was segmented into individual words, creating a sequence of tokens. This step transforms the review text into a format suitable for the LLM's processing.
    \item \textbf{Stop Word Removal}: Common words with minimal semantic meaning, such as articles (``the," ``a", ``an") and prepositions (``of," ``to"), were removed. While these words contribute to overall grammar, they often hold little value in identifying the core insights of a review.
\end{itemize}

\subsection{LLM Fine-Tuning and Model Training}
In this study, each LLM underwent a fine-tuning process using the preprocessed dataset. This process involved supervised learning, where we presented all three LLMs with labeled reviews.

To prepare the LLMs for training, we divided our preprocessed app review dataset into training and testing sets. We opted for a 70/30 split, allocating 70\% of the data for training the LLMs and 30\% for unseen testing. This split ensures the models are trained on a representative sample of the data while reserving a separate set for unbiased evaluation of their performance on reviews they have not encountered during training.

To ensure compatibility with the specific requirements of each LLM, we transformed the preprocessed review text into a format suitable for model input. We employed tokenizers explicitly designed for each model to achieve optimal compatibility between the tokenization scheme and the model architecture. For instance, BERT requires specific tokenization steps to transform the text into input tensors that the model can readily process. The \(bert\_tokenizer.tokenize\) function segments the text, and special tokens like ``[CLS]" and ``[SEP]" are added to mark the beginning and end of the sequence, respectively. Similarly, tokenizers designed for DistilBERT and GEMMA were used to prepare the text for their respective architectures. Finally, padding is applied to ensure all sequences have the same length (set to 128) using the model's specific padding token ID.

For training the LLMS, several hyperparameters were configured for the fine-tuning process. These parameters influence how the LLMs learn and adapt during training. Some of the key hyperparameters we tuned:

\begin{itemize}
    \item \textbf{Epochs}:  We opted for 5 epochs in this experiment. It is important to monitor the model's performance on the validation set (a small subset of the training data not used for direct training) to avoid overfitting. We can adjust the number of epochs based on the validation performance in future experiments.
     
    \item \textbf{Batch Size}: A batch size of 32 was chosen based on a balance between training efficiency and model performance.
     
    \item \textbf{Maximum Sequence Length}: This parameter sets a limit on the number of tokens considered within a single review during training. We chose a maximum sequence length of 128 that captures the most relevant information from the reviews while maintaining computational efficiency.
     
    \item \textbf{Learning Rate and Optimizer}: The learning rate controls the step size taken by the LLM when updating its internal weights during training. We employed an optimizer like AdamW to guide these weight updates based on the errors encountered during training with a learning rate of 2e-5.
\end{itemize}

To enhance computational efficiency and optimize memory utilization during training, we configured the model with \textit{BitsAndBytes} quantization. This technique allows the model to operate using reduced precision, thereby accelerating training without compromising performance. Additionally, we employed the Lora (low-rank) configuration, leveraging low-rank approximation techniques to streamline model training and improve scalability.

\subsection{Model Evaluation}
\label{section:mod_eval}

Having fine-tuned the LLMs BERT, DistilBERT, and GEMMA, we conducted an evaluation process to assess their effectiveness in identifying useful reviews for app developers. This evaluation serves a critical purpose, allowing us to gauge each LLM's overall performance and identify its specific strengths and weaknesses in the context of review classification.

\textbf{Evaluation Metrics}: To gain an understanding of the LLMs' performance, we employed a diverse set of metrics commonly used in text classification tasks:

\begin{itemize}
    \item \textbf{Accuracy}: It reflects the overall percentage of reviews each LLM correctly classified as ``useful" or ``not useful." While a high accuracy score is desirable, it does not necessarily tell the whole story. We dive deeper into precision and recall for a more nuanced understanding.

    \item \textbf{Precision}: It focuses on the proportion of reviews classified as ``useful" that were actually valuable for developers. A high precision score indicates the LLM effectively avoids false positives, meaning it does not mistakenly label generic reviews as useful. This is crucial, as irrelevant reviews can waste developer time and resources.
    
    \item \textbf{Recall}: Recall sheds light on how well the LLMs capture all the relevant feedback for app improvement. It measures the proportion of actual ``useful" reviews that were correctly identified by the model. A high recall score signifies the LLM's ability to comprehensively extract valuable insights from the reviews, ensuring developers do not miss important feedback.
    
    \item \textbf{F1-Score}: The F1-Score provides a balanced view by calculating the harmonic mean of precision and recall. This metric offers a more nuanced understanding of the model's performance, considering both its ability to avoid mistakes (precision) and capture all the relevant information (recall). An F1-Score closer to 1 indicates a well-performing model that excels in both aspects.
\end{itemize}

By analyzing these metrics for each LLM, we can identify the model that achieves the optimal balance between accuracy, precision and recall in the context of app review classification. This evaluation process allows us to make an informed decision regarding LLM deployment, ensuring the chosen model serves as a valuable tool for app developers.

\section{Result Analysis}

In our study, we evaluated the performance of three different language models - BERT, DistilBERT, and Gemma - using performance metrics, which are discussed in section~\ref{section:mod_eval}. Table~\ref{performance_eval} depicts the performance comparison among the LLMs model. BERT achieved the highest accuracy at 92.40\%, closely followed by Gemma at 92.00\% and DistilBERT at 91.00\%. Looking at precision, which measures how many of the predicted positive outcomes were actually correct, BERT scored the highest at 92.45\%, followed by Gemma at 92.00\%, and DistilBERT at 91.25\%. In terms of recall, which indicates how many of the actual positive outcomes were correctly predicted, Gemma led with 93.39\%, while BERT and DistilBERT both scored 92.39\% and 91.00\%, respectively. Lastly, F1-Score, a combination of precision and recall, showed BERT and Gemma tying at 92.39\% and DistilBERT close behind at 91.25\%. These metrics provide a comprehensive view of the performance of each model across different aspects of classification accuracy.

In the context of requirement elicitation, our study marks a pioneering endeavor. Given the novelty of employing LLMs in this domain, we encountered a distinct challenge in providing a comparative analysis with established or state-of-the-art methods. Consequently, our study stands at the forefront of innovation in requirement elicitation methodologies, laying the foundation for future research to explore and refine the application of LLMs in this critical area of software engineering. While lacking direct comparisons with existing methods, our pioneering work provides a crucial starting point for the integration of advanced natural language processing techniques into the realm of requirement elicitation, potentially unlocking new areas for enhancing efficiency and effectiveness in software development processes.

\begin{table}[htbp]
\caption{Performance Metrics of Language Models}
\begin{center}
\begin{tabular}{|c|p{1cm}|p{1cm}|p{1cm}|p{1cm}|}
\hline
\textbf{Models} & \textbf{Accuracy (\%)} & \textbf{Precision (\%)} & \textbf{Recall (\%)} & \textbf{F1-Score (\%)} \\ 
\hline
BERT & 92.40 & 92.45 & 92.39 & 92.39 \\ 
\hline
DISTILBERT & 91.00 & 91.25 & 91.00 & 91.25 \\ 
\hline
GEMMA & 92.00 & 92.00 & 93.39 & 91.39 \\ 
\hline
\end{tabular}
\label{performance_eval}
\end{center}
\end{table}

\begin{figure}[h]
\includegraphics[width=8cm]{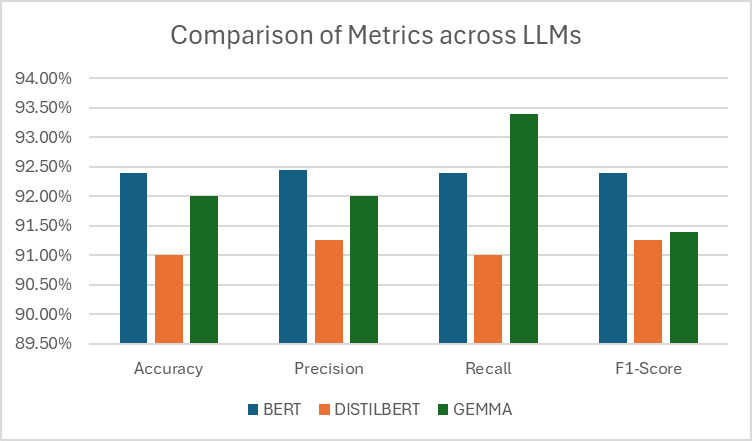}
\caption{Performance Comparison}
\label{perform_com}
\end{figure}

Fig~\ref{perform_com} presents a visual representation of the comparative performance of the various models employed in our investigation. Notably, the performance exhibited by each model appears nearly indistinguishable. However, it is imperative to underscore that the dataset utilized in our study is characterized by a relatively modest scale, which poses a substantial challenge in fine-tuning LLM architectures. Indeed, the size of the dataset constitutes a pivotal factor influencing the efficacy of model training and subsequent performance. It is conceivable that with an expansion in dataset size, the performance metrics across the models will change. This phenomenon underscores the critical importance of dataset size in the context of LLM model training and evaluation.

\section{Threat to Validity}
While this project has shown promise for using LLMs to classify app reviews, it's important to acknowledge potential threats to the validity of our findings:

\subsection{Internal Validity}

\textbf{Selection Bias:} The composition of the app review dataset used for training and testing the LLMs could influence the results. If the dataset is not representative of the real-world distribution of app reviews (e.g., skewed towards positive or negative reviews), the models might perform poorly on unseen data.

\textbf{Confounding Variables:} Other characteristics of the review or reviewer, besides the review text itself, might influence the ``usefulness" label assigned to reviews. These factors, if not accounted for, could lead to misleading conclusions about the effectiveness of the LLMs. For instance, a review's length, rating score, or the reviewer's location or demographics could be correlated with its perceived usefulness. Failing to account for these factors during model training and evaluation could lead the models to prioritize these characteristics over the actual content of the review text.

\subsection{External Validity}

\textbf{Generalizability:} The findings might not generalize well to app reviews written in different languages. LLMs used in this research are trained on massive amounts of text data in the English language, and the language used in app reviews can vary significantly across geographic regions and cultures. This variation can influence the LLM's ability to understand the nuances of the review text and make accurate classifications.

\textbf{Real-World Implementation:} The controlled environment of the research setting might not fully translate to real-world app development workflows. Factors like the user interface design and integration with existing developer tools could influence the practical effectiveness of the LLM-based solution.

\section{Discussion and Conclusion}
This research investigated the effectiveness of fine-tuned LLMs BERT, DistilBERT, and GEMMA in empowering app developers through user review analysis. Our evaluation showed that BERT has the best overall precision (the ability to identify truly valuable reviews and minimize the inclusion of irrelevant ones) and accuracy. While GEMMA demonstrated a comparatively low performance, it excelled in recall, suggesting its potential for scenarios where maximizing captured insights is crucial. These findings underscore the potential for LLMs to streamline review analysis for developers, ultimately leading to improved app quality and user satisfaction. 

To further empower developers, we see several exciting avenues for future exploration. First, collaborative research with developer communities can guide the refinement of LLM solutions to better address their specific needs. Second, designing intuitive user interfaces with clear visualizations and summaries can enhance how developers utilize the insights gleaned from LLMs.  Moreover, enhancing the dataset size and cross-validation of manually annotated reviews among the experts can be a good area to discover. Furthermore, enabling real-time review monitoring allows for proactive issue resolution and fosters stronger user trust. Finally, expanding the analysis to encompass reviews in multiple languages through multilingual LLMs can broaden the reach and impact of this technology, supporting global app development efforts. By pursuing these initiatives, we can create even more powerful LLM-based tools, solidifying their role in empowering developers to build successful and user-centric applications.

\bibliographystyle{IEEEtran}
\bibliography{refs}

\end{document}